\documentclass[aps,prl,twocolumn,groupedaddress]{revtex4}
\usepackage[pdftex]{graphicx,epsfig}

\begin{document}

\title{Anomalously slow attrition times for asymmetric populations \\ with internal group dynamics}

\author
{Zhenyuan Zhao$^{1}$, Juan Camilo Bohorquez$^{2}$, Alex Dixon$^3$ \& Neil F. Johnson$^{1}$}
\affiliation{$^{1}$Department of Physics, University of Miami, Coral Gables, FL 33126, U.S.A.\\
$^{2}$Department of Industrial Engineering, Universidad de Los Andes, Bogota, Colombia\\
$^{3}$Department of Physics, Clarendon Laboratory, Cambridge University, Cambridge CB3 0HE, U.K.}

\date{\today}

\begin{abstract} The many-body dynamics exhibited by living objects include group formation within a population, and the non-equilibrium process of attrition between two opposing populations due to competition or conflict. We show analytically and numerically that the combination of these two dynamical processes generates an attrition duration $T$ whose nonlinear dependence on population asymmetry $x$ is in stark contrast to standard mass-action theories. A minority population experiences a longer survival time than two equally balanced populations, irrespective of whether the majority population adopts such internal grouping or not. Adding a third population with pre-defined group sizes allows $T(x)$ to be tailored. Our findings compare favorably to real-world observations.
\end{abstract}

\maketitle
Predator-prey systems have been widely studied by many disciplines, including physics\cite{McKane}. Outside the few-particle limit, mean-field mass action equations such as Lotka-Volterra provide a reasonable description of the average and steady-state behavior, i.e. $dN_A(t)/dt=f(N_A(t),N_B(t))$ and $dN_B(t)/dt=g(N_A(t),N_B(t))$ where $N_A(t)$ and $N_B(t)$ are the $A$ and $B$ population at time $t$. Such population-level descriptions of living systems do not explicitly account for the well-known phenomenon of intra-population group (e.g. cluster) formation\cite{caro}, leading to intense debate concerning the best choice of functional response terms for $f(N_A(t),N_B(t))$ and $g(N_A(t),N_B(t))$ in order to partially mimic such effects\cite{cosner}. Analogous mass-action equations have been used to model the interesting non-equilibrium process of attrition (i.e. reduction in population size) as a result of competition or conflict between two predator populations in colonies of ants, chimpanzees, birds, Internet worms, commercial companies and humans\cite{nicola} in the absence of replenishment. The term attrition just means that `beaten' objects become inert (i.e. they stop being involved) {\em not} that they are necessarily destroyed. 
The combined effects of intra-population grouping dynamics and inter-population attrition dynamics have received suprisingly little attention\cite{caro,redner1}, despite the fact that grouping and attrition are so widespread\cite{nicola,caro} and the fact that their coexisting dynamics generate an intriguing non-equilibrium many body problem.

In this Letter, we consider explicitly the effect of intra-population grouping dynamics on the duration $T$ of attrition between two opposing populations $A$ and $B$. In stark contrast to the standard mass-action theories, we find that $T$ exhibits a maximum for highly asymmetric predator-predator systems in the absence of population replenishment (Fig. 1(a)). This non-monotonic dependence of $T$ on population asymmetry is remarkably insensitive to whether the majority (i.e. larger) population exhibits internal grouping or not (Fig. 1(b)). We show how $T$ can be manipulated by the addition of a third population which blocks encounters involving smaller fragments. Within the physics community, Redner and co-workers had considered a related problem of conflict within a clustering population in one dimension, and highlighted intriguing connections to a more general class of roughening phenomena in physics\cite{redner1}. Eguiluz and Zimmerman and others had considered an infinite-range one population coalescence-fragmentation model as a simplified version of human opinion formation\cite{ez}, while Galam and others have considered interesting models involving competition and conflict\cite{opinion,wattis}. Our work focuses instead on the consequences of coexisting grouping and attrition on the duration $T$ (i.e. the survival time of the minority population).

The intra-population group dynamics in our model are driven by essentially the same coalescence-fragmentation processes as Ref. \cite{ez}, while the inter-population attrition process is essentially the same as Ref. \cite{redner1} (Fig. 1(c)). We have checked that our main conclusions are robust to a variety of reasonable generalizations (e.g. randomly selecting groups independent of group size, or attrition beyond a simple cluster subtraction rule\cite{redner1}, or allowing for a limited number of new recruits over time) and to a reasonably wide range of parameter space. Our model combines the following specific mechanisms:
It is well documented that groups of objects (e.g. animals, people) may suddenly scatter in all directions (i.e. complete fragmentation) when its members sense danger, simply out of fear\cite{caro} or in order to confuse a predator\cite{caro}. (Curiously, clusters of inanimate objects such as doubly-ionized Argon atoms and animal Hox genes, also exhibit such complete fragmentation\cite{Argon}). Since a sense of danger can arise at any time, our model randomly selects a candidate group for fragmentation at each timestep, with probability proportional to its size since larger groups have more members and hence are increasingly likely to spot danger or be themselves spotted\cite{caro}. With probability $\nu_A$  or $\nu_B$ for groups of type $A$ or $B$, the group fragments completely. If it doesn't fragment, a second group is randomly selected with probability again proportional to size, since any subsequent coalescence and attrition events will likely be initiated by pairwise interaction between individual members in the two groups and hence the probability will depend on the number of members.  If the group is of the same type (i.e. $A$ or $B$) the two groups coalesce, mimicking the observation that groups may try to build up their size to increase their security\cite{caro}. `Coalescence' can simply mean that two groups act in a coordinated way, not necessarily that they are physically joined. If of opposite type, their interaction leaves an $A$ or $B$ group of size $|s_A-s_B|$ if $s_A>s_B$ or $s_B>s_A$ respectively (or zero is $s_A=s_B$) where $s_A$ and $s_B$ are $A$ and $B$ group sizes (Fig. 1(c)). Other forms of attrition rule (e.g. stochastic) can yield similar results. At time $t$, populations $A$ and $B$ hence comprise $n^A_{s}(t)$ and $n^B_{s}(t)$ groups (i.e. clusters)
of size $s$, where $\sum sn^A_{s}(t)=N_{A}(t)$ and $\sum sn^B_{s}(t)=N_{B}(t)$. 
Interactions are distance-independent as in Ref. \cite{ez} since we are interested in systems where messages can be transmitted over arbitrary distances (e.g. modern human communications). Bird calls and chimpanzee interactions in complex tree canope structures can also mimic this setup, as may the increasingly longer-range awareness that arises in larger animal, fish, bird and insect groups\cite{caro}. 

\begin{figure}
\noindent
\includegraphics[width=0.49\textwidth]{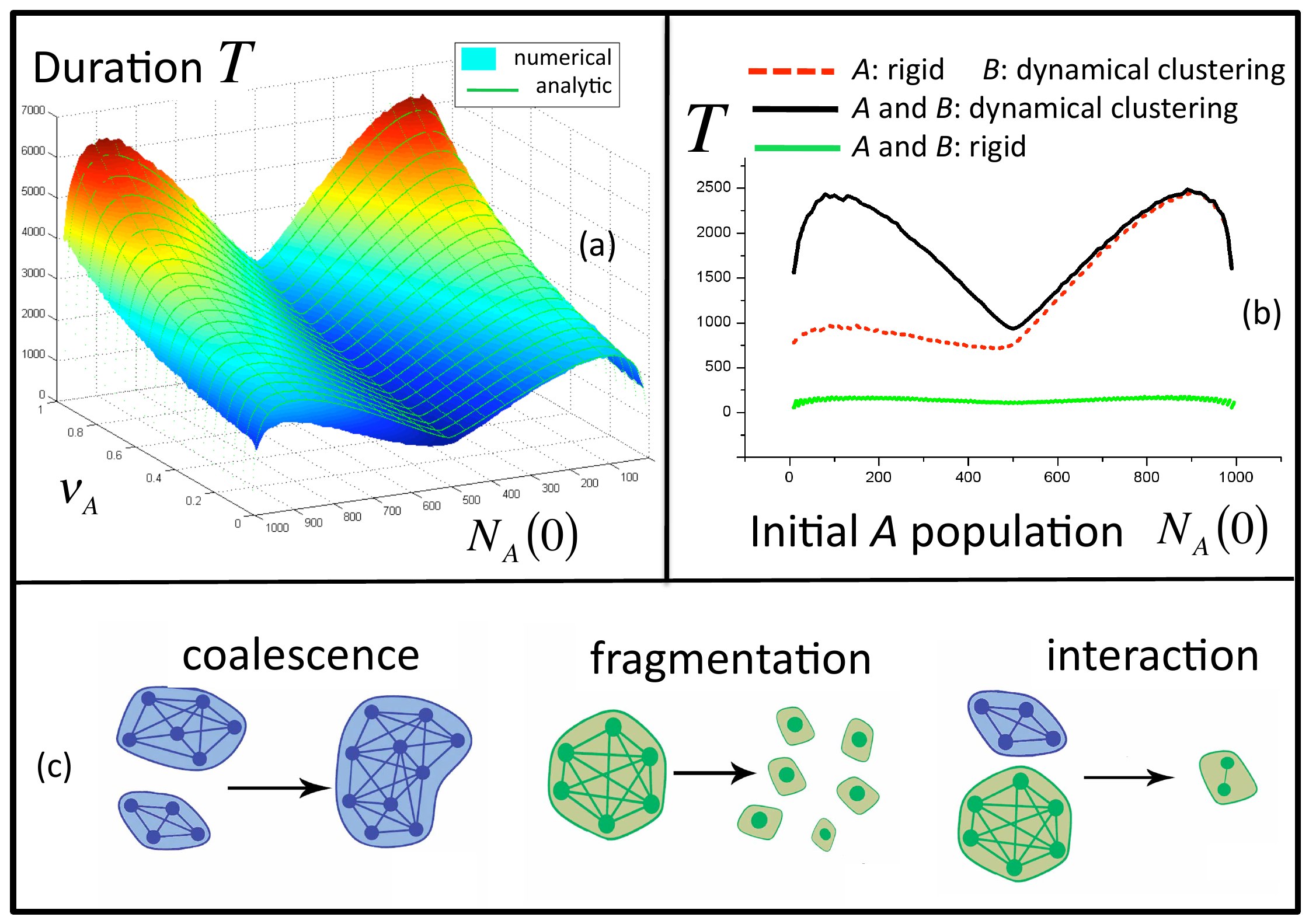}
\caption{\label{fig:extinction3}
(a) Duration $T$ of attrition (or equivalently, the extinction time or survival time of the smaller population)
as a function of initial $A$ population $N_A(0)$ and fragmentation probability $\nu_A$ ($\nu_{B}=0.3$). Solid lines are analytic (Eq. (2)) while the surface is numerical simulation. $N_A(0)+N_B(0)=1000$.
Qualitative features are unchanged by varying  $\nu_{B}$.  (b) Black curve: same as Fig. 1(a) with $\nu_{A}=\nu_{B}=0.3$. Red dashed curve: $A$ contains rigid units (size 10) while $B$ features internal dynamical grouping (i.e. clustering) as explained in text. Green curve: both $A$ and $B$ comprise rigid units (size 10).
(c) Events in our model. Two groups of same type can coalesce, e.g. $6+4=10$. Individual groups can fragment, e.g. $6\rightarrow 6\times 1$. Two groups of opposite type interact, e.g. group of size $6-4=2$ remains.}
\end{figure}

Figure 1(a) shows our numerical and analytic results (Eq. (2)) for the duration $T$. The initial condition for the numerical simulations comprises isolated individuals, however the curve is insensitive to these initial conditions since the initial group (i.e. cluster) formation times act like a small additive term. The present two-population numerical implementation is a straightforward generalization of the one-population version discussed in Ref. \cite{ez}. The excellent agreement  suggests that our analytic treatment of internal grouping using a time-averaged interaction term may have wider application within non-equilibrium many body systems. Following the model mechanisms discussed above, the probability $Q_{AB}$ that any $A$ cluster is selected and
interacts with a $B$ cluster is the sum over all $s$ of the
probability for an $A$ cluster of size \(s\) to interact with any
$B$ cluster, which gives $(1-\nu_{A}) {N_{A}(t)
N_{B}(t)}/[N_{A}(t)+N_{B}(t)]^{2}$. The probability $Q_{BA}$ is 
similar, with $\nu_{A}$ replaced by $\nu_{B}$. After an
interaction, $A$ and $B$ are reduced by the size of the
smallest interacting cluster, whose average value $c$ (i.e. average interaction size) is well approximated by unity plus a
small linear correction term
$0.2(1+x)^{-1}(1-x)(1-\nu_A)(1-\nu_B)$ because clusters are generally very small over a large proportion of $T$. Note $c$ is not formally the same as the average cluster size -- in part because interactions do not occur at every timestep and the entire system is actually time-dependent -- but they tend to take on
similar values. Employing constant $c$,
the populations after $i$ interactions become
$N_A(t)=N_{A}(0)-ic$, $N_B(t)=N_{B}(0)-ic$. The probability
for an interaction between $A$ and $B$ clusters after \(i\)
previous interactions is $Q(i)= Q_{AB} + Q_{BA}$ and hence
\begin{equation}
Q(i) =  \frac{{(N_A(0) - ic)(N_B(0) - ic)}}{{(N_A(0) + N_B(0) - 2ic)^2}}(2 - \nu_A -\nu_B)
\end{equation}
To reduce $N_A(t)$ and $N_B(t)$ by \(c\) takes \(1/Q(i)\) timesteps on average. The time to reduce one population to zero is the sum of the timesteps required for each interaction, until the population is eliminated.
Supposing $B$ is the smaller population, it requires \(N_B(0)/c\) interactions to eliminate it, hence the final interaction happens after \(N_B(0)/c - 1\) previous interactions. The time $T$ to eliminate the smaller population $B$ is therefore $\sum{{Q^{-1}(i)}}$ with $i$ running from $0$ to $\frac{{N_B(0)}}{c} - 1$.
Using  \(\sum_{1}^{n}{\frac{1}{i}} = \gamma + \psi_0(n+1)\), where \(\gamma\) is the Euler-Mascheroni constant and \(\psi_0\) is the digamma function, and \(\sum_{a+1}^{n} = \sum_{1}^{n} - \sum_{1}^{a}\), we obtain the duration:
\begin{eqnarray}
\small
\small
\label{eqn:extinction}
T &=& \frac{N_A(0)-N_B(0)}{c(2-\nu_{A}-\nu_{B})}
\Big[\frac{4N_B(0)}{N_A(0)-N_B(0)}\nonumber \\
& & +\big[\gamma+\psi_{0}\Big(\frac{N_B(0)}{c}+1\Big)\big]\\
& &-\big[\psi_{0}\Big(\frac{N_A(0)}{c}+1\Big)-\psi_{0}\Big(\frac{N_A(0)-N_B(0)}{c}+1\Big)\big]
\Big]\ \ . \nonumber
\end{eqnarray}
When $A$ is the smaller population, the form is identical
but with $A$ and $B$ interchanged. This $T$ expression 
depends only on the initial populations of $A$ and $B$, hence $T
\equiv T(x,\{\nu\})$ for constant $N$, where $\{\nu\}\equiv
(\nu_A,\nu_B)$. Differentiation yields a maximum $T$ at $x_{\rm
max} \simeq 0.788$ for $\nu_A=\nu_B$ and $N=10^3$, independent of
$\{\nu\}$. Numerical simulations confirm that $T(x,\{\nu\}) \sim
t_1(\{\nu\})t_2(x)$, with $t_1(\{\nu\}) \sim 1/(2-\nu_A-\nu_B)$ exactly as in Eq. (2), thereby supporting our use of a constant average interaction size $c$. 
Two factors therefore determine the duration $T$: one
originates from the grouping dynamics within a given
population (i.e. $t_1(\{\nu\})$), while the other originates from
the asymmetry (i.e. $t_2(x)$). Replacing sums by
integrals, Eq. (2) can be approximated in the
large population limit as:
\begin{eqnarray}
T_{\rm cont} &=& \frac{N_A(0)-N_B(0)}{c(2-\nu_{A}-\nu_{B})}
\Big[{\rm ln}\frac{N_B(0)[N_A(0)-N_B(0)+c]}{c N_A(0)}\nonumber \\
& & + 4\Big(\frac{N_B(0)-c}{N_A(0)-N_B(0)}
\Big)
\Big] \ .
\end{eqnarray}
The peak at $x_{\rm max}$ is robust to a variety of model variants, and can be understood as follows: When $x\sim 0$, clusters of $A$ and $B$ are abundant and have a reasonably large average size. Interactions between $A$ and $B$ clusters are frequent and the attrition per interaction is high, hence $T$ is small.
As $x$ increases, with $A$ being the larger population, an interaction between an $A$ and $B$ cluster is increasingly likely to eliminate the $B$ cluster completely since the $A$ cluster is increasingly likely to be the larger cluster. However the interaction rate is decreasing rapidly, and $T$ increases overall. For $x\rightarrow 1$, it may take a long time to find a $B$ cluster however there are very few to find, hence $T$ becomes smaller.  Interestingly, the distribution of time-intervals between interactions of $A$ and $B$ clusters is approximately exponential for all $x$, except near $x_{\rm max}$ where it becomes approximately power-law.
Note that if the attrition were to end after a given fraction of the initial population is eliminated, the same qualitative results would still emerge since the theory is essentially invariant under overall changes of timescale. 
Figure 1(b) shows the results of $A$ and/or $B$ adopting different internal grouping. The duration $T$ remains essentially unchanged if the larger population chooses a static internal structure comprising rigid units of a particular size. If the smaller population adopts such rigid units, $T$ decreases significantly. Hence $T$ is largely dictated by the internal group dynamics of the minority population. If both $A$ and $B$ are internally rigid, $T$ is small for all $x$.

\begin{figure}
\includegraphics[width=0.48\textwidth]{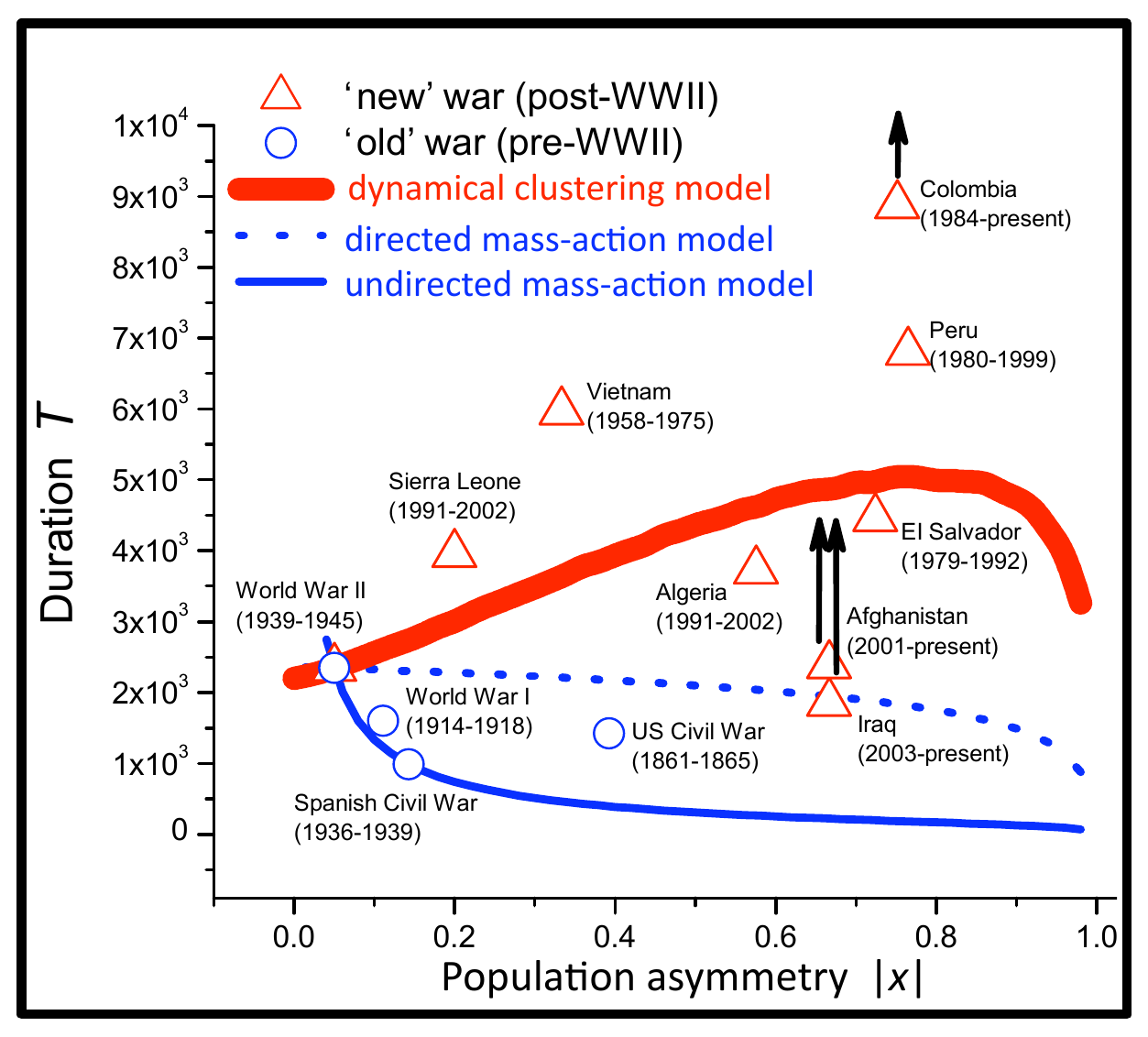}
\caption{Duration $T$ of human conflicts, as function of asymmetry $x$ between the two opposing military populations. $x=|N_A(0)-N_B(0)|/[N_A(0)+N_B(0)]$. Data are up to the end of 2008, hence final datapoints for the three ongoing wars will lie above positions shown, as indicated by arrows. Lower two blue lines are the mass-action results.
Upper red curve (i.e. Eq. (2)) generated using  $\nu_{A}=\nu_{B}=0.7$ and $[N_A(0)+N_B(0)]$ fixed (as in (a)).  Changing  
$\nu_{A}$ and $\nu_{B}$ changes height of theoretical peak, but leaves qualitative features unchanged.}
\end{figure}

\begin{figure}
\includegraphics[width=0.48\textwidth]{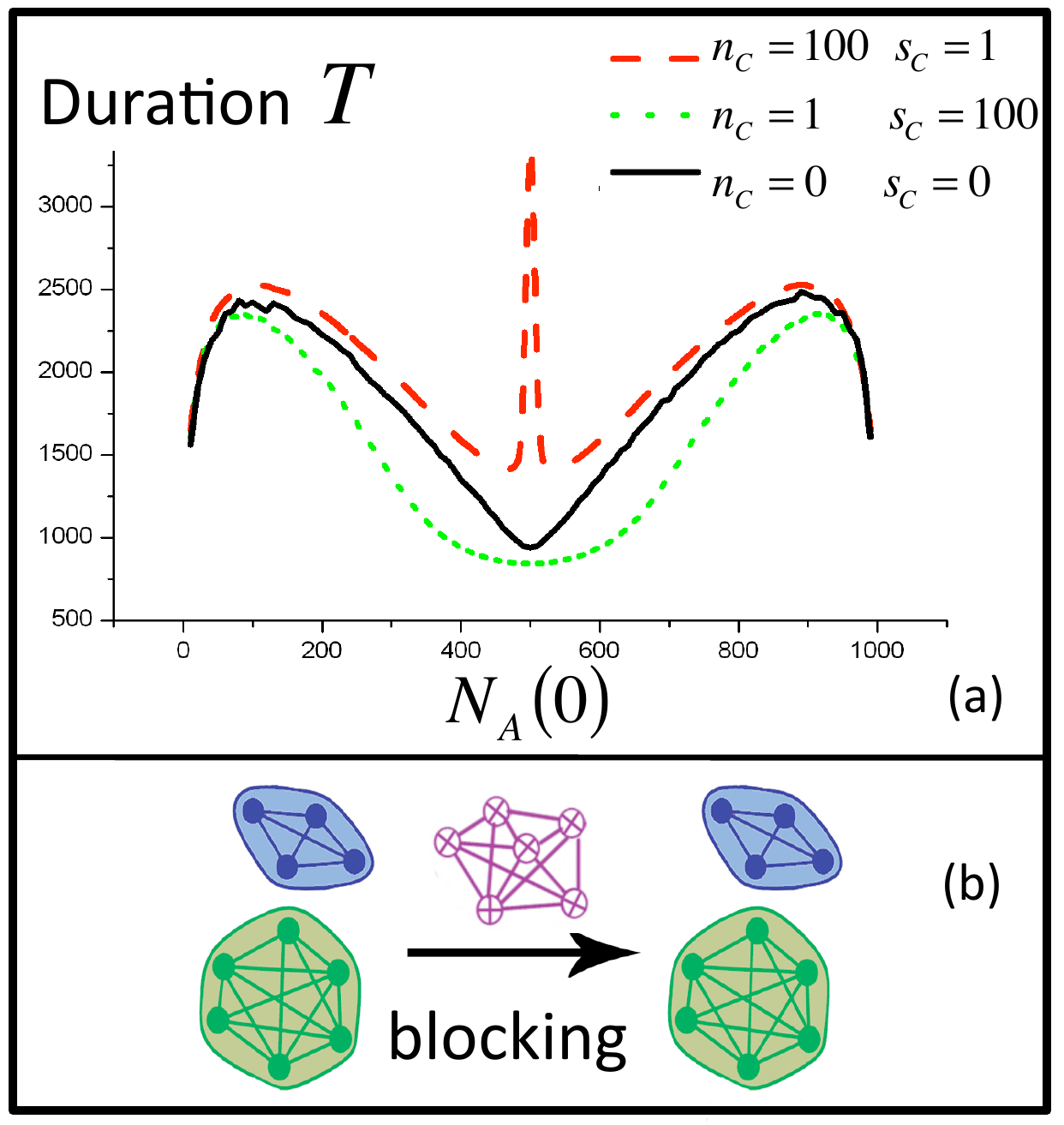}
\caption{\label{fig:variations}
(a) Black curve as in Figs. 1(a), 1(b), 2(b) with $A$ and $B$ undergoing internal
dynamical clustering. $\nu_A=\nu_{B}=0.3$. Red dashed curve: $n_C=100$ third-party groups, each of size $s_C=1$. Green dotted curve: $n_C=1$ third-party group, of size $s_C=100$.
(b) Third-party blocking event. If neither
$A$ or $B$ clusters are bigger than $C$ cluster, then $C$ cluster blocks the interaction and permanently neutralizes both clusters.}
\end{figure}

The top red curve in Fig. 2 compares our theory to empirical results for human conflicts, while the lower two blue curves show the mass-action predictions. The mass-action equations traditionally used for attrition are\cite{nicola}: (1)
$d N_A(t)/dt=-u_L
N_A(t) N_B(t)$ and $d N_B(t)/dt=-u_L N_A(t) N_B(t)$, called Lanchester's undirected mass-action model; (2) $ d N_A(t)/dt=-d_L N_B(t)$ and $d N_B(t)/dt=-d_L
N_A(t)$, called Lanchester's directed mass-action model\cite{nicola} where $u_L$ and $d_L$ are constants. 
`Old' wars are blue circles and `new' wars are red triangles, with World War II labelled by both since it is a natural dividing point. Since $N\gg 1$, we take the end-point for the undirected mass-action model to correspond to reducing the smaller population to one instead of zero, thereby avoiding problems with a continuum description of $N_A(t)$ and $N_B(t)$ near zero.
Figure 2(b) offers some support for a recent hypothesis in the social science domain, distinguishing between old wars in which $A$ and $B$ adopt traditional, fairly rigid, military structures, and new wars in which $B$ (and possibly $A$) adopt more fluid tactics akin to our model\cite{kaldor}.  By contrast, the `old' wars are well described by both the green curve of Fig. 1(b) (i.e. rigid armies) and the traditional mass-action theories (blue curves), implying that such internal group dynamics were absent in `old' wars. 

Figure 3(a) shows that the duration $T$ can be manipulated by adding a third-party
population $C$ which can block interactions (Fig. 3(b)). For simplicity, we assume the  $N_C$ members of $C$ are permanently arranged into $n_C$ groups
each with $s_C$ permanent members. Apart from peacekeepers in human conflict, $C$ could mimic the targeted blocking of interactions between particular physiological clusters. $A$ and $B$ undergo dynamical clustering as before, except that if a $C$ group is selected and it is bigger or equal to the size of the
$A$ and $B$ clusters, the interaction is blocked and the two $A$ and $B$ clusters are permanently pacified (i.e. neutralized). Figure 2(a) shows that if $C$ comprises only a few, large groups (e.g. green dotted curve) then $T$ decreases irrespective
of the asymmetry. Having a few, large $C$ groups means that some sizeable battles can be blocked, however it also allows the build-up of sizeable groups of both $A$ and $B$ which in turn makes the typical size of interactions bigger. By contrast, if $C$
comprises many small groups (e.g. red dashed curve) $T$ can
be much larger, showing a huge increase around $x\sim 0$. If real-time management of the $C$ population is possible, this duration profile can be manipulated even further.

We stress that our model findings are not a simple consequence of either dilution leading to reaction slow-down, or of the specific cluster selection scheme that we chose. In our model, as in nature, opposing predator groups actively seek each other out at each timestep, even if their density is low, making this unlike simple chemical dilution, and hence unlike simple mass-action equations. Regarding cluster selection, we have verified numerically that our main conclusions are unchanged if we select clusters independent of size, or use other fragmentation schemes (e.g. binary splitting into two clusters). This is because the smaller population spends the majority of the conflict as very small groups or individuals, hence the weighting by size is not so important. In short, our results emerge from the interplay between population asymmetry, the presence of clustering, and the intentional engagement between the two opposing populations. Although the specific consequences may vary by application area, we believe that related phenomena lying beyond mass-action predictions will arise in a wide range of physical, chemical, biological and social systems, whenever intra-population clustering coexists with inter-population reactions.
  
We thank R. Denney and M. Spagat for very useful discussions.

\end{document}